\documentclass[showpacs, pra,onecolumn,preprintnumbers ,amsmath, amssymb, superscriptaddress, aps]{revtex4-2}
\usepackage{graphicx}
\graphicspath{{Plots/}}
\usepackage{dcolumn}
\usepackage{bm}
\usepackage[colorlinks]{hyperref}
\usepackage{physics}
\usepackage{physconst}
\usepackage{physunits}
\usepackage{caption}
\usepackage{subcaption}
\usepackage{epstopdf}
\usepackage{nameref}
\usepackage{float}

\def\b{\begin{equation}}
\def\e{\end{equation}}

\begin{document}
	
	
	\title{
		Quantum Hall coherent perfect absorption in graphene}
	
\author {Dariush Jahani\footnote{\href{dariush110@gmial.com}{dariush110@gmial.com}}}
\author{Mohammadreza Alikhani}
\author{Yaser Abdi }

	\affiliation {Department of Physics. University of Tehran, Tehran, Iran	
	}

	\date{\today}
	
	\begin{abstract}
\textbf{\abstractname.} $100\%$ absorption in a two-dimensional electron gas (2DEG) with Dirac spectrum is demonstrated to be obtained by controlling the interference of multiple incident radiations, referred to as coherent perfect absorption (CPA). However, when a 2DEG such as graphene is exposed to a magnetostatic bias, it resonantly could absorb electromagnetic radiation by transitions of its Dirac electrons between non-equidistant and nonlinear Landau levels. Here, the magneto-optical terahertz (THz) CPA in graphene under the quantum Hall effect (QHE) regime at both strong and subtesla magnetostatic bias fields is addressed. Our findings show that an effective magneto-optical surface conductivity corresponding to right- and left-handed circular (RHC- and LHC) polarizations could model a magneto-tunable CPA in graphene in THz range. Significantly, graphene under QHE regime reveals different tunable CPA properties for each circularly polarized beams by the intensity of the applied magnetic bias. Moreover, it is observed that different phase modulations at CPA frequencies are achieved for RHC and LHC polarizations. Considering the maximum efficiency for a 2D absorber, our results demonstrate the magnetostatic tuning of CPA in 2D Dirac materials for long-wavelength sensing applications and signal processing.
	\end{abstract}
	\pacs{}
	\maketitle

	\section{Introduction}
Under a normally applied magnetic field, unlike conventional 2D electron gas (2DEG) systems, the Landau levels (LLs) in 2D Dirac materials are not equally spaced and vary non-linearly with the field modulations. This could significantly affect the absorption properties of 2DEG systems with Dirac spectrum. Graphene, a 2D material composed of carbon atoms arranged in a honeycomb pattern, has attracted huge attention due to its exceptional electronic and optical properties \cite{1,2}. Interestingly, the non-equidistant and nonlinear properties of LLs of graphene in the presence of a magnetic field which make it a promising material for various optoelectronic applications arises from its linear energy dispersion relation \cite{3,4}. It is well-known that the zero-gap Dirac cones in the linear energy spectrum of graphene give rise to observation of the unconventional quantum Hall conductivity in the presence of a magnetic bias applied normally to its surface\cite{5}.

Now, the absorption of light is being studied to understand how electromagnetic waves can be harvested to stimulate electrons in materials. For instance, in optical sensors and solar cells \cite{6, 7, 8, 9, 10}, the primary phenomenon is the absorption of incoming electromagnetic waves through materials and converting them to electrical current. The absorption of materials could be considered due to the electronic excitations in near-infrared and visible, phonons' local vibrational modes in mid-infrared, and collective modes in terahertz regime \cite{11}. However, the absorption of electromagnetic waves in a atomically 2D materials is generally insufficient for real-world applications. Therefore, developing innovative methods to enhance the absorption of them is an essential step toward improving its optical absorption properties.

 From a theoretical standpoint, the maximum absorption of a 2D material is $50\%$ of the incident radiation. Up to know, several methods have been proposed to increase the absorption of 2D materials, such as fabry-perot cavities \cite{12, 13, 14, 15, 16} to trap waves in an optical cavity and excitation surface plasmonic resonances \cite{17, 18, 19}. Now, since these approaches are highly dependent on the structure, absorption is typically limited to a narrow electromagnetic spectrum. As a result, additional techniques are needed to achieve perfect absorption in a more straightforward and tunable manner. Here, coherent perfect absorption (CPA) is a promising concept to address these challenges. The underlying principle behind CPA is making standing waves by interfering with two beams with tuned frequency to trap electromagnetic energy onto the absorber and surpassing the scattering matrix. Therefore, the perfect absorption could be accessible by tuning the phase modulation of two incident beams \cite{20, 21}. Graphene, as a one-atom-thick 2D material, in the visible and near-IR zone under normal incidence could absorb only $2.3\%$ of light. While this is a relatively high value for an atomically thin layer, it is still far too insufficient from being a perfect absorber of light.

Recently, graphene has been studied for its potential to achieve coherent perfect absorption (CPA) in the mid-infrared and terahertz ranges. Shraddha M. Rao et al. showed 80\% modulation of light through the CPA concept. They employed unstructured multilayer graphene as an absorber in their setup and employed a continuous wave laser source (532 nm, 500 $\mu$ W) to produce coherent illumination.  In another approach, an absorber consisting of 30 layers of graphene was sandwiched between two silica substrates to create resonances in the absorber \cite{22}. Yuancheng Fan et al. calculated the CPA frequency required to maintain perfect absorption. They considered monolayer graphene and illuminated it with two coherent beams.  By finding the appropriate values for $|r|$ and $|t|$ to suppress the scattering matrix, they obtained a CPA frequency of 1.735 THz for doped monolayer graphene ($E_{F} = 0.5$ eV).  Moreover, by tuning the phases of the two beams, they found the optimal phase to enhance the absorption to $100\%$ (at a phase modulation equal to $0.435\pi$) \cite{23}.

 Up to now, the coherent absorption in graphene can be tuned substantially by varying the gate-controlled Fermi energy level which control the optical transitions of graphene.  Optical conductivity of graphene is also a function of the illuminating frequency and the electron relaxation time. Some efforts have been made to change the Fermi level of graphene through chemical doping or applying voltage \cite{24, 25, 26, 27}. However, applying a magnetostatic bias field could be another way to manipulate the graphene's optical transitions between nonlinear LLs under quantum Hall effect (QHE) situation. To be more, specific, applying a magnetic field to the surface of graphene can introduce nonlinear Landau energy levels in its linear spectrum. This leads to a Hall term in the tensor of its magneto-optical conductivity which could significantly alter CPA conditions in QHE regime. In this work, for the first time, we propose that introducing a magnetic bias field into the unstructured monolayer graphene results in a magneto-tunable CPA at both sub-tesla and high field levels for RHC and LHC polarizations in THz regime. Our results could pave the way for magnetically tunable perfect absorber devices and ultrafast graphene-based quantum magnetometers.
	
\section{Model and theory }\label{Theo}
	
	Our model is schematically proposed in figure 1 for a single-layer graphene sheet under an applied magnetic bias. Here, graphene is perpendicularly illuminated along $z$-direction by two circularly-polarized coherent counter-propagating beams, $I^{(1)}_{\pm}$ and $I^{(2)}_{\pm}$ with the output magnitudes of $O^{(1)}_{\pm}$ and $O^{(2)}_{\pm}$, respectively. It is well known that the effect of the constant magnetic field on the graphene layer is quantizing its energy spectrum in a nonlinear way as $E_{n} =\pm\sqrt{2nv_{F}^{2}|eB|h}$  with $n=0,1,2, ...$ to indicates LLs. This magneto-quantizing effect leads to an antisymmetric surface conductivity tensor for graphene with longitudinal ($\sigma_{L}$) and Hall conductivity ($\sigma_{H}$) terms which could be coupled in an effective magneto-optical conductivity  $\sigma_{\pm}=\sigma_{L}\pm i\sigma_{H}$ for RHC and LHC polarizations \cite{28}. Both longitudinal and Hall conductivity terms could be obtained by the use of the Kubo formula:\cite{3}:
\begin{equation}
\begin{split}
\sigma_{L}(\omega)=
\frac{e^{2}v_{f}^{2}\vert eB\vert\left(\hbar\omega+2i\Gamma \right)}{\pi i}\times\\ \sum_{n=0}^{\infty}\left\lbrace\frac{\left[f_{d}(M_{n})-f_{d}(M_{n+1})\right]+ \left[f_{d}(-M_{n+1})-f_{d}(-M_{n})\right]}{\left(M_{n+1}-M_{n}\right)^{3}-\left(\hbar\omega+2i\Gamma \right)^{2}\left(M_{n+1}-M_{n}\right)}\right\rbrace \\
+ \left\lbrace\frac{\left[f_{d}(-M_{n})-f_{d}(M_{n+1})\right]+ \left[f_{d}(-M_{n+1})-f_{d}(M_{n})\right]}{\left(M_{n+1}+M_{n}\right)^{3}-\left(\hbar\omega+2i\Gamma \right)^{2}\left(M_{n+1}+M_{n}\right)}\right\rbrace\\
\end{split}
\end{equation}
and
\begin{equation}
\begin{split}
\sigma_{H}(\omega)=\frac{-e^{2}v_{f}^{2} eB}{\pi}\times
\sum_{n=0}^{\infty}\left\lbrace \left[f_{d}(M_{n})-f_{d}(M_{n+1})\right]- \left[f_{d}(-M_{n+1})-f_{d}(-M_{n})\right]\right\rbrace \times \\
\left\lbrace \frac{1}{\left(M_{n+1}-M_{n}\right)^{2}-\left(\hbar\omega+2i\Gamma \right)^{2}}+\frac{1}{\left(M_{n+1}+M_{n}\right)^{2}-\left(\hbar\omega+2i\Gamma \right)^{2}}\right\rbrace
\end{split}
\end{equation}
where $f_{d}(\varepsilon)=\frac{1}{1+exp( \frac{\varepsilon-\mu_{c}}{K_{B}T})}$ and $K_{B}$ is the Boltzmann constant. The scattering coefficients of two outputs. i.e, $O^{(1)}_{\pm}$ and $O^{(2)}_{\pm}$ could be related to two RHC and LHC polarized input beams  $I^{(1)}_{\pm}$ and $I^{(2)}_{\pm}$ by a scattering matrix $S_{\pm}$ as:
\begin{equation}
\begin{bmatrix}
O^{(1)}_{\pm}\\
O^{(2)}_{\pm}
\end{bmatrix}
=S_{\pm}
\begin{bmatrix}
I^{(1)}_{\pm}e^{i\phi^{(1)}_{\pm}}\\
I^{(2)}_{\pm}e^{i\phi^{(2)}_{\pm}}
\end{bmatrix}=\begin{bmatrix}
t^{(1)}_{\pm}&&r^{(2)}_{\pm}\\
t^{(1)}_{\pm}&&r^{(2)}_{\pm}
\end{bmatrix}
\begin{bmatrix}
I^{(1)}_{\pm}e^{i\phi^{(1)}_{\pm}}\\
I^{(2)}_{\pm}e^{i\phi^{(2)}_{\pm}}
\end{bmatrix}
\end{equation}		
Here, due to the reciprocity and spatial symmetry of monolayer graphene, we put $t^{(1)/(2)}_{\pm}=t_{\pm}$ and $r^{(1)/(2)}_{\pm}=r_{\pm}$. To proceed, we introduce transmission and reflection coefficients via an effective transfer matrix corresponding to the normal propagation of RHC and LHC polarized light in graphene in QHE regime. Considering the surface conductivity tensor of graphene written in the following matrix form:
\begin{equation}
\sigma
=\begin{bmatrix}
\sigma_{xx}&& \sigma_{xy}\\
-\sigma_{xy}&& \sigma_{xx}
\end{bmatrix}
\end{equation}	
and also the relations $\sigma_{xx}=\sigma_{yy}=\sigma_{L}$ and $\sigma_{xy}=-\sigma_{yx}=\sigma_{H}$, one can show that the scattering coefficients for RHC and LHC polarizations could be obtained by the following equation:
\begin{equation}
\begin{bmatrix}
t_{1\pm}\\
r_{1\pm}
\end{bmatrix}
=\frac{1}{2}\begin{bmatrix}
2+z_{0}\sigma_{\mp}&& z_{0}\sigma_{\mp}\\
-z_{0}\sigma_{\mp}&& 2-z_{0}\sigma_{\mp}
\end{bmatrix}
\begin{bmatrix}
t_{2\pm}\\
r_{2\pm}
\end{bmatrix}
\end{equation}	
 Now, from this transfer matrix formalism one could prove that, transmission and reflection coefficients in the presence of a constant magnetic field for RHC and LHC polarizations are:
	\begin{equation}
t_{\pm}=\frac{2}{2 +z_{0}(\sigma_{L}\mp i\sigma_{H})}
\end{equation}
\begin{equation}
r_{\pm}=-\frac{z_{0}(\sigma_{L}\mp i\sigma_{H})}{2 +z_{0}(\sigma_{L}\mp i\sigma_{H})}
\end{equation}
where $z_{0}$ is the impedance of the free space. Note that RHC and LHC polarizations result in different transmission and reflection coefficients for graphene at normal incidence. This is because electronic transitions of non-interacting Dirac fermions between Landau energy levels are different. The transitions for RHC and LHC polarizations, $-n+1\rightarrow n$ and $-n\rightarrow n-1$ produce different responses in the magneto-optical conductivity of graphene. Therefore, the phase modulation of the coherent input circularly polarized beams could be obtained by the following relation:
\begin{equation}
|O^{(1)}_{\pm}|=|O^{(2)}_{\pm}|=| t_{\pm}I_{\pm}e^{i\phi^{(1)}_{\pm}}+ r_{\pm}I_{\pm}e^{i\phi^{(2)}_{\pm}}|
\end{equation}
from which the necessary condition for CPA performance is $|t_{\pm}|=|r_{\pm}|$. Therefore, the necessary condition for quantum Hall magneto-CPA is $|\sigma_{\pm}|=2/z_{0}$. It is obvious for $\sigma_{H}=0$ above relation yields equal results for RHC and LHC polarizations and recover the former CPA condition in the absence of a magnetic bias. It is straightforward to show that the magneto-absorption at CPA frequency could be written as follows:
\begin{equation}
A_{\pm}(\omega_{CPA})=1-4|t_{\pm}|^{2}I_{\pm}^{2}[1+\cos(\Delta_{\pm}(\omega_{CPA})+\Delta\phi_{\pm})]
\end{equation}
where $\Delta_{\pm}=\varphi_{t_{\pm}}-\varphi_{r_{\pm}}$ and $\Delta\phi_{\pm}=\phi^{(1)}_{\pm}-\phi^{(2)}_{\pm}$. From the above expression, it is clear that, for fixed magnetometric bias applied on graphene, the only thing that plays the central role in achieving a perfect absorbtion is the phase modulation of the two input beams, $\Delta\phi_{\pm}$. In fact, $100\%$ absorption for a 2DEG with a Dirac spectrum under a magnetic bias is demonstrated to be obtained by controlling the interference of RHC and LHC polarized radiations such that:
\begin{equation}
\Delta\phi_{\pm}=n\pi-\Delta_{\pm}(\omega_{CPA})
\end{equation}
It should be emphasized that for a non-zero $\Delta_{\pm}$, the phase-difference for RHC and LHC polarized beams depends on the magneto-CPA frequency and, therefore, could be generaly different for each circularly polarizations.
	
\section{Results and discussion}

In order to investigate the effect of magnetic field on magneto-CPA properties of graphene, we first calculate the transmission and reflection intensities for two sets of magnetostatic bias fields. As it is shown in figure \ref{Figure_2a}, and figure \ref{Figure_3a}, Te and Re are calculated for LHC and RHC polarized modes passing normally through a graphene layer under an applied external constant magnetic field. Although the chemical potential and temperature can directly affect graphene's effective magneto-optical conductivity, as mentioned in section \ref{Theo}, we set $\mu=0.2\  eV $ at both low temperature($10 \ K$) and the room temperature ($T=300 \ K$) where the most notable changes is shown to occur. The Fermi velocity is consider to be $v_{F}=3\times 10^{6} ms^{-1}$. In this work the many-body effects which causes the Fermi velocity of massless carriers of graphene to be different for electronic transitions between LLs is not considered. Also, we ignore the dependence of $\Gamma$ (which is considered to be $0.11 \ meV$ in our simulations) on Landau energies.

Considering LHC polarized modes, as shown in figure \ref{Figure_2b}, at sub-tesla magnetic fields for $T=10\ K$, one notices that CPA frequencies show up at $0.94$ and $1.18$ THz which corresponds to applied sub-tesla magnetic fields $B=0.3$ and $B=0.6$ Tesla, respectively. Whereas at strong fields the magnitude of the effective magneto-optical surface conductivity of monolayer graphene intersects the line $2/z_{0}$ at more points. Equivalently, transmission and reflection curves could show intersections at two or even more different CPA frequencies when the magnetic field is $\ge 1$ Tesla. More precisely, at $B=1$ Tesla (figure \ref{Figure_2a}) CPA condition occurs at $0.09$ and $1.49$ THz. Note that, as it is depicted in figure \ref{Figure_2b}, the maximum absorption condition occurs at frequencies where $z_0 |\sigma| = 2$. We see that, in general, Left-handed CPA frequencies are blue-shifted by increasing the magnetic field bias. In fact, under the magnetic bias $B=4$ Tesla, CPA frequencies in graphene for LHC polarizations increase to 2.50 and 3.90 THz. Interestingly, applying larger magnetic field gives a resonant behavior for effective magneto-optical surface conductivity so that at $B=6$ Tesla, one can observe four CPA frequencies appearing at 3.91, 4.94, 4.99, and 5.45 THz. Surprisingly, as illustrated in figures \ref{Figure_2c} and \ref{Figure_2d} at $T=10\ K$, at applying magnetic bias fields more than one Tesla there is no CPA frequency for RHC polarizations. However, at sub-tesla fields $B=0.3$ and $B=0.6$ Tesla, we have magneto-CPA frequencies for RHC modes occurring at $0.46$ and $0.22$ THz at $T=10\ K$, respectively (figure \ref{Figure_2d}). This is contrary to behaviour of the magneto-CPA properties of graphene for LHC polarizations for which strong magnetic fields blue-shifted the CPA points. It seems that unlike LHC modes, applying an applied stronger magnetic bias to graphene red-shifts the right-handed CPA frequency. 

For the magnetic bias below $B=1$ Tesla, magneto-CPA frequencies are shown as a function of the chemical potential in figure \ref{Figure_3a}. The effect of increasing the chemical potential for both RHC and LHC polarizations on CPA points show that coherent absorption occurs at higher frequencies. Now, the effect of increasing the temperature to $T=300\ K$ on the magneto-CPA properties of graphene in QHE situation is depicted in figure \ref{Figure_3b}. This figure shows that the number of magneto-CPA frequencies for LHC polarizations is similar to case when the temperature is $T=10 \ K$. However, the frequencies appear at different points, 0.04 and 1.53 THz at one Tesla. Here, magneto-CPA frequencies for LHC polarized modes appear at $0.94$ and $1.19$ THz for $B=0.3$ and $B=0.6$ Tesla, respectively, which almost do not show changes in comparison to the low temperature case. The significant results here is that stronger magnetostatic bias blue-shifts LHC polarized CPA frequencies. The reason is that at stronger magnetic fields photons with more energy are needed to excite electronic transitions between well-separated LLs. Moreover, at the room temperature, when the magnetic field increases, the resonances of effective magneto-optical conductivity significantly increase for LHC polarizations. More precisely, there exists 10, 18, and 12 CPA frequencies corresponding to the $B=4$, $B=5$, and $B=6$ Tesla, respectively. In figure \ref{Figure_3c} the magneto-CPA frequencies for RHC and LHC polarizations as a function of the magnetic bias are depicted. One can notice that at stronger magnetic fields RHC and LHC magneto-CPA frequencies are linearly red- and blue-shifted with more separated space, respectively. As it is expected at zero magnetic field the optical response of graphene for both circular polarizations is the same and, therefore, RHC and LHC CPA frequencies are seen to merge to the same frequency.  In fact, in the absence of an applied magnetic bias since Hall conductivity is equal to zero effective magneto-optical surface conductivities $\sigma_{\pm}=\sigma_{L}\mp i\sigma_{H}$ for RHC and LHC polarizations get similar value $\sigma_{\pm}=\sigma_{g}$. At the room temperature $T=300\ K$, magneto-CPA frequencies appear at $0.46$ and $0.24$ THz for sub-tesla fields $B=0.3$ and $B=0.6$ Tesla, respectively (figure \ref{Figure_3d}). Therefore, similar to LHC modes temperature does not affect RHC polarized CPA frequencies.

The absorption of two LHC polarized coherent electromagnetic beams is shown in figure \ref{Figure_4}. The coherent absorbtion is measured as a function of the phase difference and frequency of the two input beams illuminating graphene under different applied magnetic fields at the low temperature limit. As it is observed, the proper phase modulation of two input beams plays a central role in the absorption spectra of graphene for a fixed bias. By tuning the proper phase of the input beams, as implied by the equation (9), the system can either absorb $100\%$ of the incoming waves at CPA points or let them pass through completely. Thus, optimizing the beam's phases for each magneto-CPA frequency is necessary to achieve 100\% absorption. In figure \ref{Figure_4a} it is depicted that $99.99\%$ absorption occurs for an applied bias $B=1$ Tesla at 0.09 THz ( $\Delta\phi=0.477 \pi$) and 1.49 THz ( $\Delta\phi=1.525 \pi$ ). As it is shown in figures \ref{Figure_4b}, \ref{Figure_4c} and \ref{Figure_4d}, at magnetic bias fields $B=4$, $5$, and $6$ Tesla, the proper phase modulation of the two input beams remain the same. However, at $B=6$ Tesla, two new CPA frequencies appear at 4.94 and 4.99 THz. Here, at CPA points, $\Delta\phi$ is obtained to be $1.710\pi$ and $1.891\pi$. In figure \ref{Figure_5}, the coherent absorbtion spectrum for graphene in QHE regime is plotted in the room temperature $T=300\ K$. At this temperature new absorbtion regions can be observed due to thermal excitations between LLs. In figures \ref{Figure_5b}, \ref{Figure_5c} and \ref{Figure_5d} the phase modulations at the room temperature for magnetic fields $B=4$, $B=5$ and $B=6$ Tesla are shown, respectively. It is clear that, still, the general trend is blue-shifting the magneto-CPA region at stronger bias fields.

Now, we turn our attention to the low field limit for highly doped graphene. As illustrated in figure \ref{Figure_6}, the magneto-absorption of monolayer graphene for the chemical potential $\mu=0.5 \ eV$ at temperature $T= 10\ K$ under $B=0.3$ Tesla and $B=0.6$ Tesla is evaluated. For LHC polarization, magneto-CPA points in the frequency space appear at 1.859 and 1.954 THz with an optimized phase modulation of two beams at $\Delta\phi=1.509 \pi$ for $B=0.3$ and $B=0.6$ Tesla, respectively. Considering RHC polarization, for highly doped graphene the maximum absorption (99.9999\%) emerges at 1.668 and 1.572 THz with an optimized phase modulation of incident lights at $\Delta\phi=1.509\pi$. Increasing the Fermi energy give graphene much more metallic properties so that CPA-points appear at higher frequencies in comparison to case of lower Fermi energy $\mu=0.2\ eV$.

\section{CONCLUSION}

In conclusion, we prove that monolayer graphene could exhibit THz CPA in the QHE regime for circularly polarized radiations. Also, tunable property of quasi-CPA frequency corresponding to right- and left-handed modes by the intensity of the applied magnetic field was simulated. Moreover, at CPA frequencies, the two circularly polarizations revealed different phase modulations for an applied magnetic field intensity. It was demonstrated that two or even more CPA frequencies with different properties could be emerged for left-handed modes. Interestingly, it was shown that increasing the magnetic field bias for right-handed (left-handed) polaized beam red-shift (blu-shift) the CPA frequency. We also examined CPA properties at subtesla magnetic fields and found that the CPA frequency could be modulated ranging from about 1.4 to 2 THz. It was then revealed that our results at low magnetic field could recover earlier CPA conditions reported for graphene in the absence of an applied magnetostatic bias. The physics behind our results could be much more clear by analyzing $|\sigma_{\pm}|$ for THz frequency range and LLs properties of graphene under the applied magnetostatic bias on graphene. Surprisingly, the interaction of two coherent counter-propagating electromagnetic beams with nonlinear Landau energy levels in graphene demonstrated a linear modulations of magneto-CPA frequency shifts for both RHC and LHC polarizations as a function of the magnetic field bias. This linear modulation of RHC and LHC magneto-CPA frequencies was shown to be dependent on the Fremi energy. Our results could pave the way to novel approaches in designing tunable ultrathin terahertz absorbers and on-chip optical magnetometers.

\section{Data Availability} 
All data generated or analysed during this study are included in this published article [and its supplementary information files].

\section{References}
	\bibliographystyle{plain}
	
\break

\begin{figure}[H]\label{Figure_1}
  \centering
  \includegraphics[width= \textwidth]{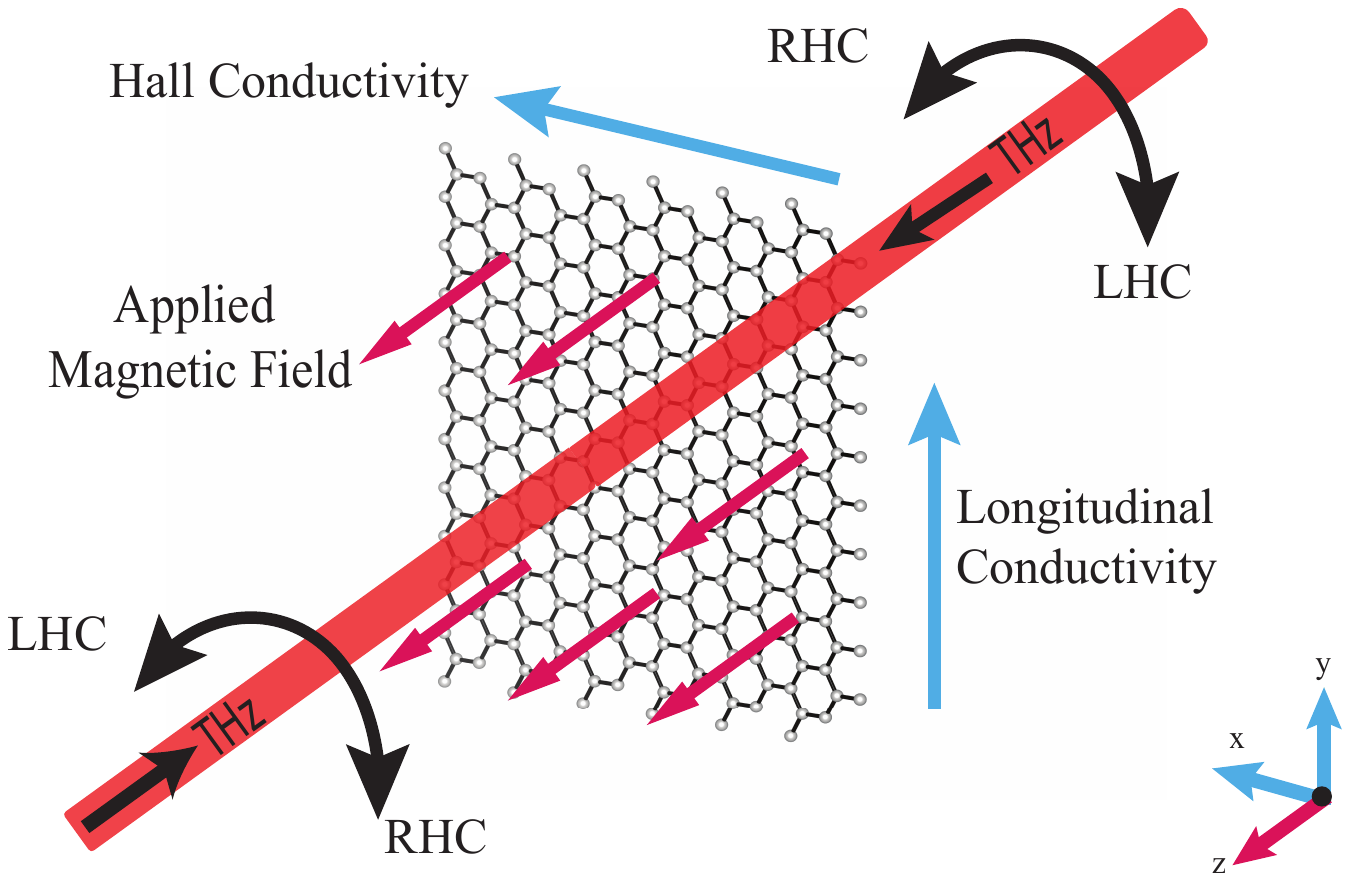}
  \caption{Schematic representation of the introduced magneto-CPA system.  Monolayer graphene is illuminated by two circularly polarized counter-propagating beams under an applied magnetic bias. The longitudinal and Hall conductivity of the system is modeled by an effective magneto-optical surface conductivity $\sigma_{\pm}=\sigma_{L}\pm i\sigma_{H}$ for graphene layer associated with RHC and LHC polarizations. }
\end{figure}
   \begin{figure}[H]
\centering
    \begin{subfigure}{.45\textwidth}
        \centering
        \includegraphics[width = \textwidth]{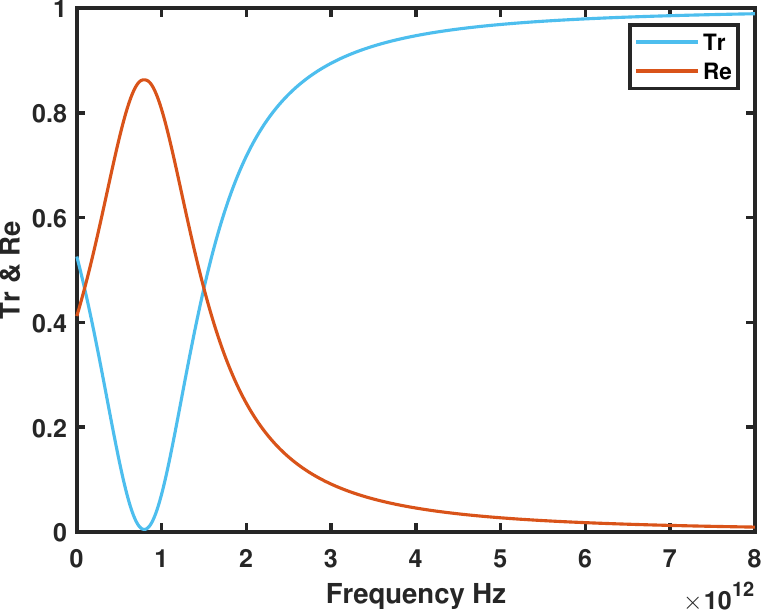}
        \caption{}
        \label{Figure_2a}
    \end{subfigure}
    \begin{subfigure}{.45\textwidth}
        \centering
        \includegraphics[width = \textwidth]{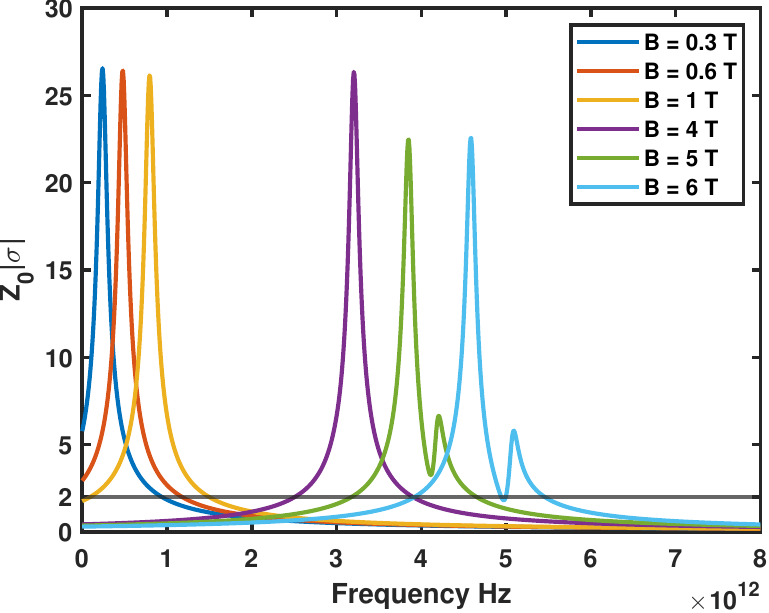}
        \caption{}
        \label{Figure_2b}
    \end{subfigure}
    \begin{subfigure}{.45\textwidth}
        \centering
        \includegraphics[width = \textwidth]{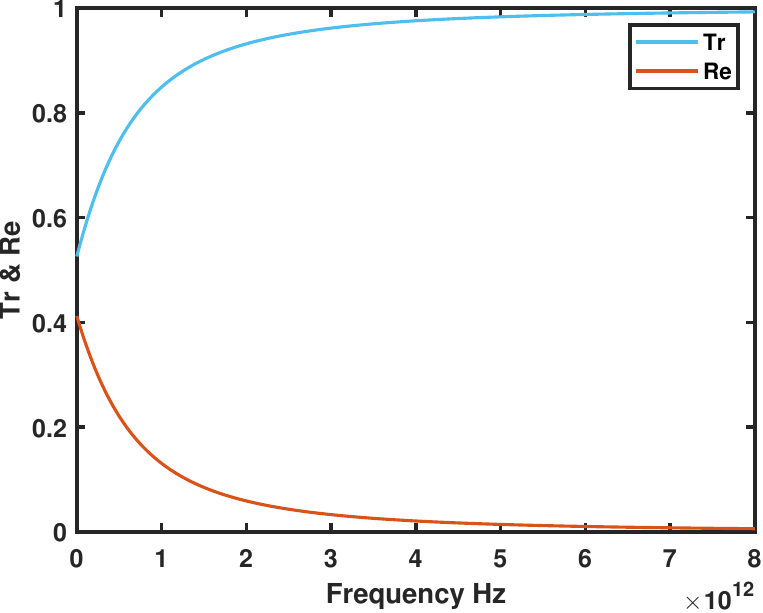}
        \caption{}
        \label{Figure_2c}
    \end{subfigure}
    \begin{subfigure}{.45\textwidth}
        \centering
        \includegraphics[width = \textwidth]{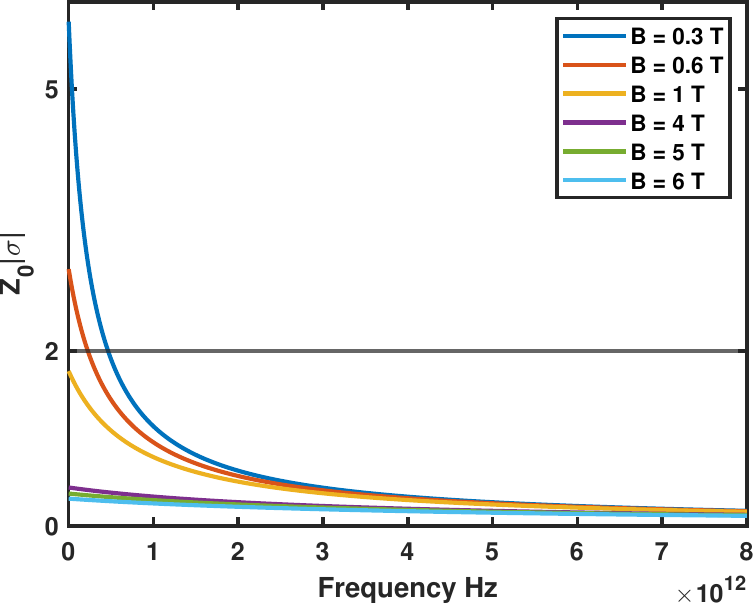}
        \caption{}
        \label{Figure_2d}
    \end{subfigure}

    \caption{(a) Transmission and reflection for applied magnetic field $B=1\ T$, chemical potential $\mu= 0.2\ eV$ and temperature $T= 10\ K$ considering LHC polarization. (b) $z_{0}|\sigma|$ as a function of the frequency of the incident beams for different magnetostatic bias fields. (c)(d) The same situations for RHC polarizations. It is seen that for LHC polarizations above sub-tesla fields CPA could occur at two or even more frequencies. Interestingly, for RHC polarizations CPA frequencies are proved to occur only at sub-tesla fields. }
    \label{Figure_2}
\end{figure}
\begin{figure}[H]
\centering
    \begin{subfigure}{.45\textwidth}
        \centering
        \includegraphics[width = \textwidth]{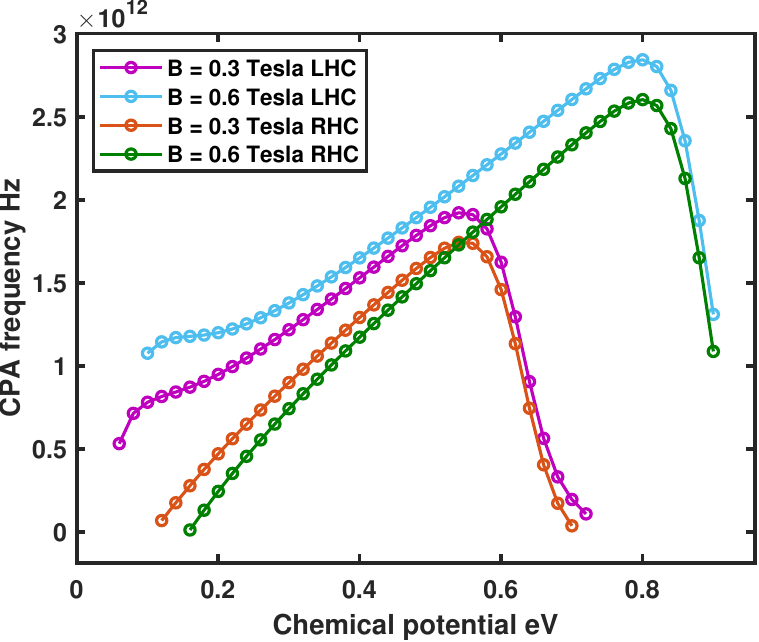}
        \caption{}
        \label{Figure_3a}
    \end{subfigure}
    \begin{subfigure}{.45\textwidth}
        \centering
        \includegraphics[width = \textwidth]{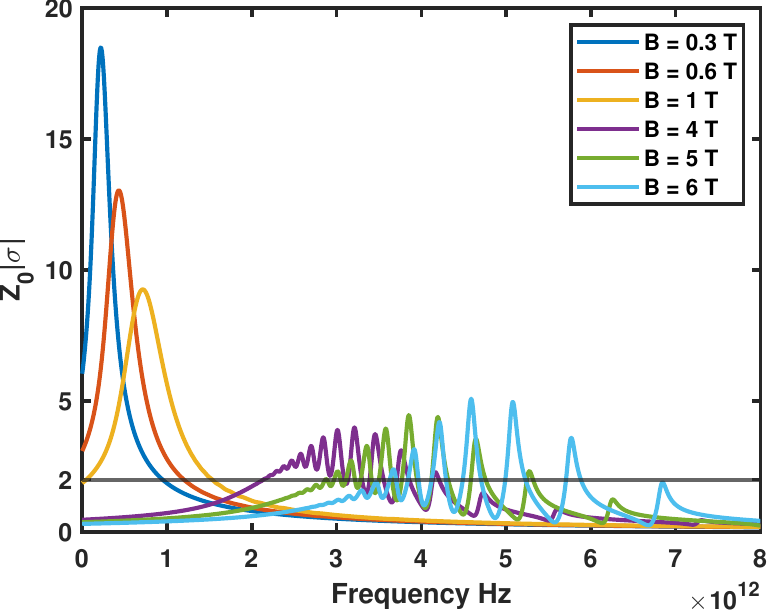}
        \caption{}
        \label{Figure_3b}
    \end{subfigure}
    \begin{subfigure}{.45\textwidth}
        \centering
        \includegraphics[width = \textwidth]{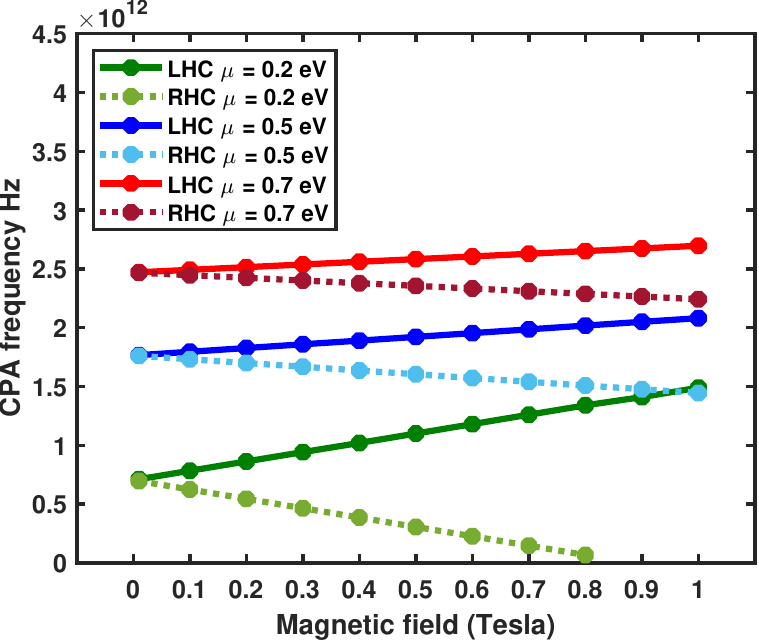}
        \caption{}
        \label{Figure_3c}
    \end{subfigure}
    \begin{subfigure}{.45\textwidth}
        \centering
        \includegraphics[width = \textwidth]{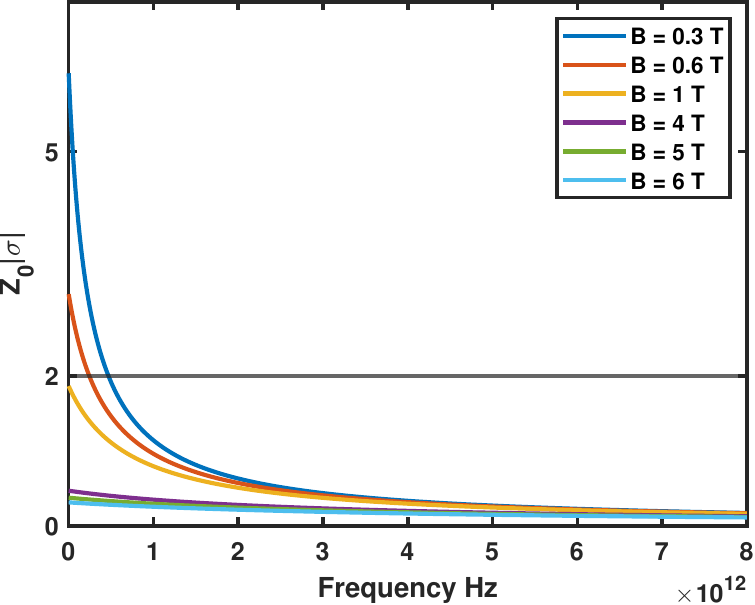}
        \caption{}
        \label{Figure_3d}
    \end{subfigure}
    \caption{(a) Magneto-CPA frequencies as a function of the chemical potential at $T=300\ K$. (b) $Z_0|\sigma|$ plot with respect to incident beams frequency at the room temperature for LHC polarization. It is observed that the rise of the temperature creates more magneto-CPA points with broader distributions at stronger fields (c) The magneto-CPA situation for RHC and LHC beams as a function of the magnetic bias. It can be seen that at stronger magnetic fields RHC and LHC magneto-CPA frequencies are more separated and merge to the same frequency at zero bias (d) $Z_0|\sigma|$ as function of the frequency for RHC modes. Magneto-CPA points are observed to emerge only at sub-tesla fields. It is clear that effective magneto-optical surface conductivity is diminished at stronger bias fields for RHC modes.}
    \label{Figure_3}
\end{figure}
\begin{figure}[H]
\centering
    \begin{subfigure}{.45\textwidth}
        \centering
        \includegraphics[width = \textwidth]{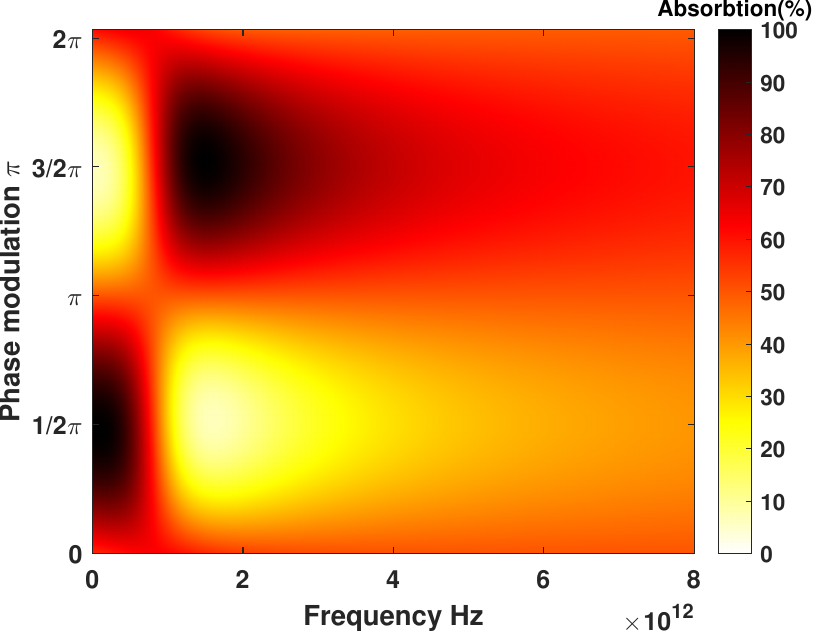}
        \caption{B = 1 Tesla.}
        \label{Figure_4a}
    \end{subfigure}
    \begin{subfigure}{.45\textwidth}
        \centering
        \includegraphics[width = \textwidth]{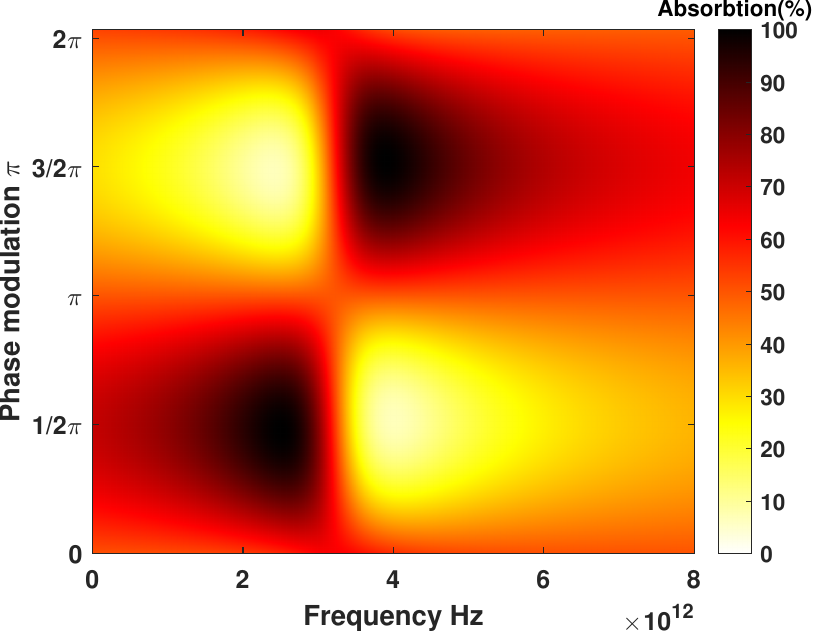}
        \caption{B = 4 Tesla.}
        \label{Figure_4b}
    \end{subfigure}
    \begin{subfigure}{.45\textwidth}
        \centering
        \includegraphics[width = \textwidth]{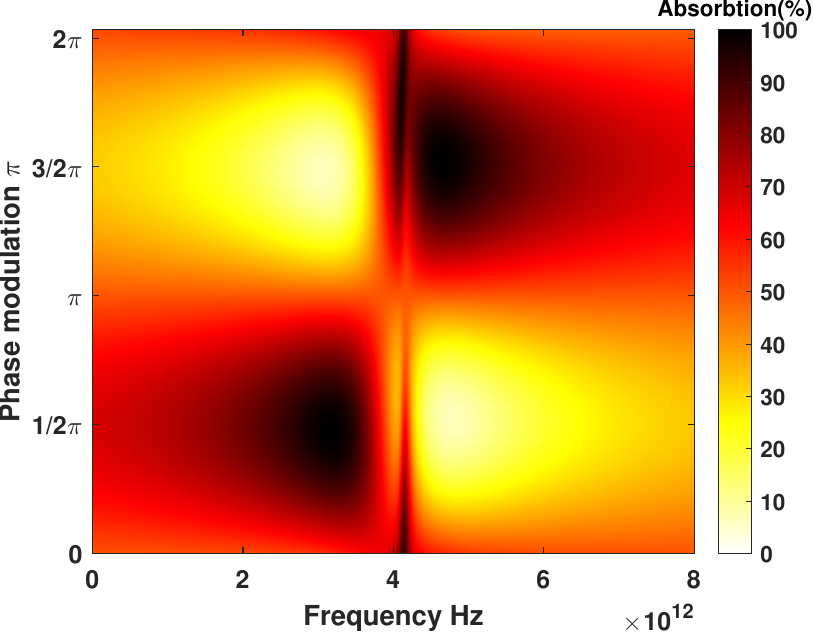}
        \caption{B = 5 Tesla.}
        \label{Figure_4c}
    \end{subfigure}
    \begin{subfigure}{.45\textwidth}
        \centering
        \includegraphics[width = \textwidth]{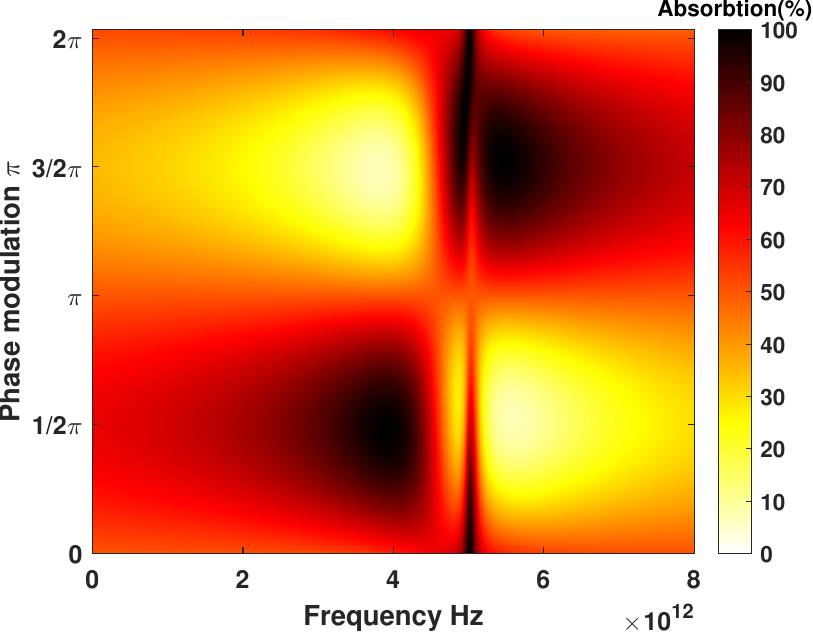}
        \caption{B = 6 Tesla.}
        \label{Figure_4d}
    \end{subfigure}
\caption{Coherent absorbtion for LHC polarized beams as a function of the frequency and phase medullation for different applied magnetic fields at the low temperature limit $T = 10\ K$. (a) There exists two broad magneto-CPA region at the bias $B=1$ Tesla (b) The presence of four broad magneto-CPA region in higher frequency zone at the bias $B=4$ Tesla. (c) Emergence of extra narrow perfect magnet-absorptions in higher CPA frequencies at $B=5$ Tesla (d) Magneto-CPA properties at $B=6$ Tesla is relatively more blue-shifted. Note that extra Magneto-CPA resonances appear in higher frequencies relative to the first magneto-CPA point.  }
\label{Figure_4}
\end{figure}
\begin{figure}[H]
\centering
    \begin{subfigure}{.45\textwidth}
        \centering
        \includegraphics[width = \textwidth]{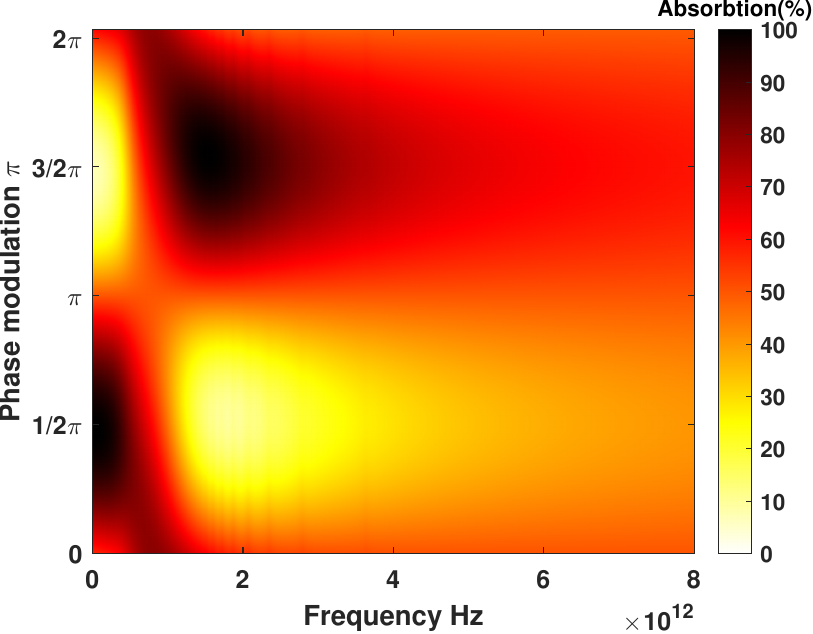}
        \caption{B = 1 Tesla.}
        \label{Figure_5a}
    \end{subfigure}
    \begin{subfigure}{.45\textwidth}
        \centering
        \includegraphics[width = \textwidth]{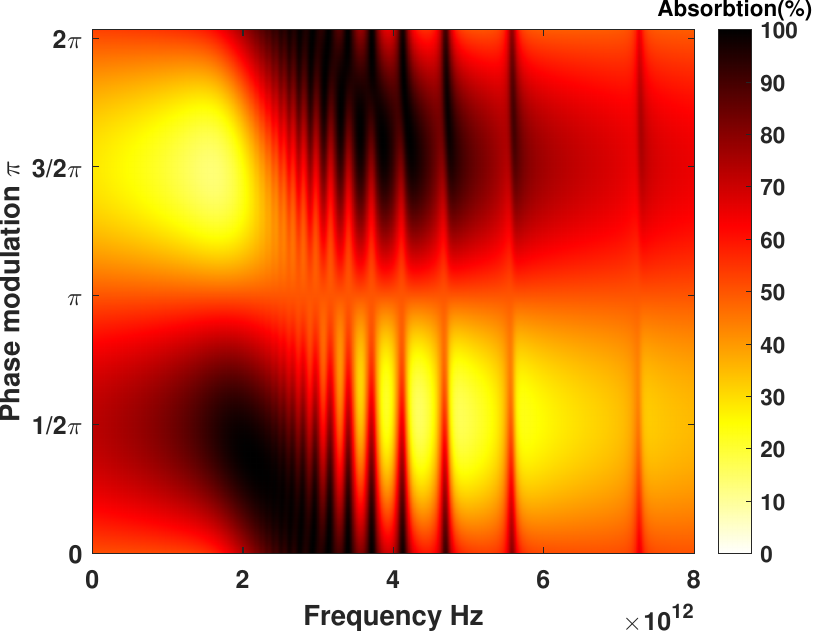}
        \caption{B = 4 Tesla.}
        \label{Figure_5b}
    \end{subfigure}
    \begin{subfigure}{.45\textwidth}
        \centering
        \includegraphics[width = \textwidth]{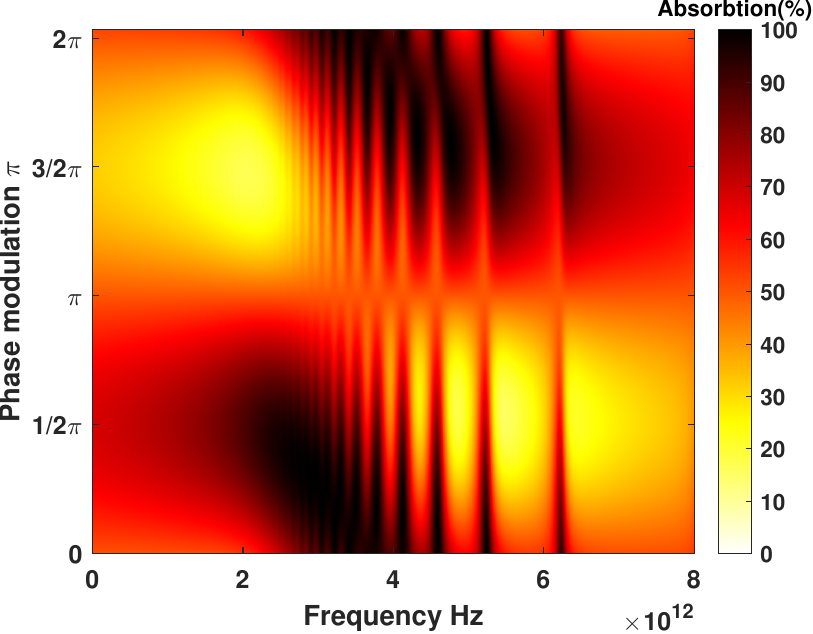}
        \caption{B = 5 Tesla.}
        \label{Figure_5c}
    \end{subfigure}
    \begin{subfigure}{.45\textwidth}
        \centering
        \includegraphics[width = \textwidth]{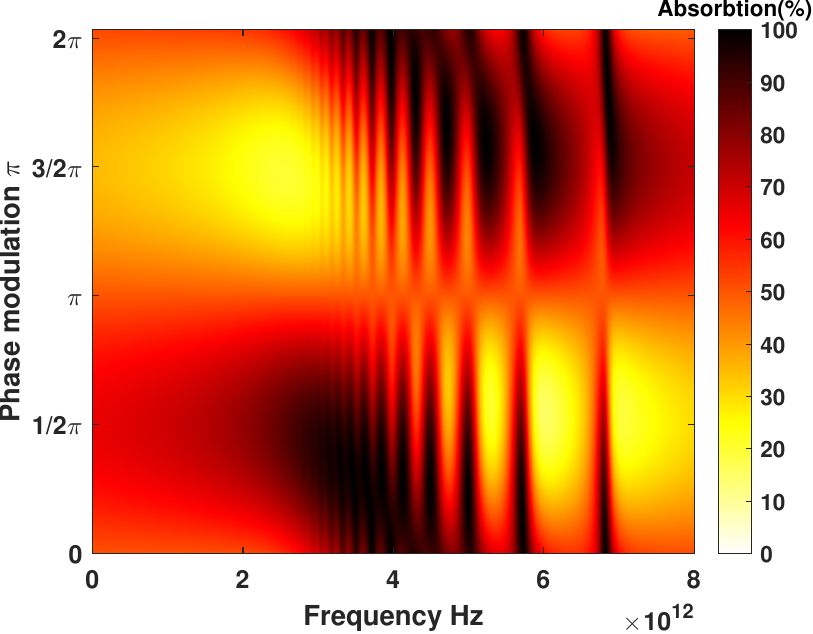}
        \caption{B = 6 Tesla.}
        \label{Figure_5d}
    \end{subfigure}
\caption{ Coherent absorbtion of LHC polarized beams as a function of the frequency and phase medullation for different applied magnetic fields at the room temperature $T = 300\ K$. It is observed that a significant rise in the temperature at applied bias fields more that one Tesla on graphene leads to emergence of more narrow-band magneto-CPA  resonances. }
\label{Figure_5}
\end{figure}

\begin{figure}[H]\label{Figure_6}
\centering
    \begin{subfigure}{.45\textwidth}
        \centering
        \includegraphics[width = \textwidth]{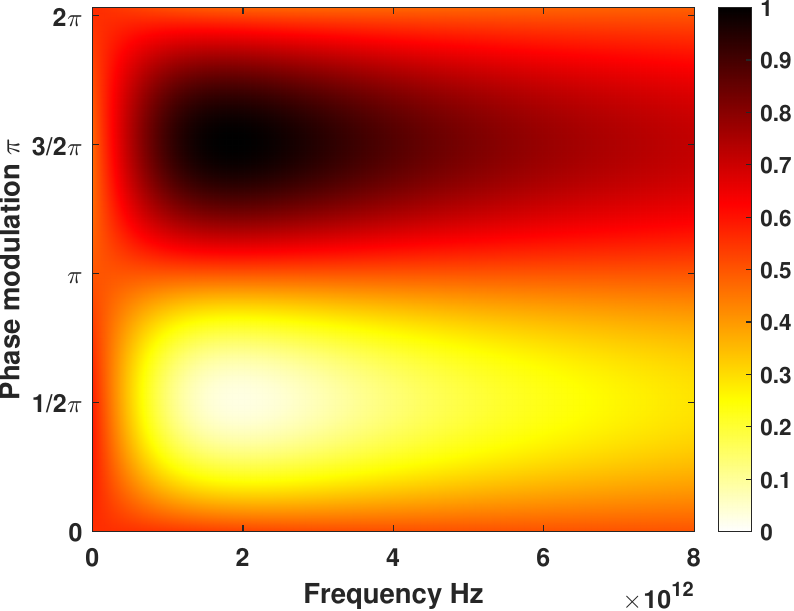}
        \caption{B = 0.3 Tesla.}
    \end{subfigure}
    \begin{subfigure}{.45\textwidth}
        \centering
        \includegraphics[width = \textwidth]{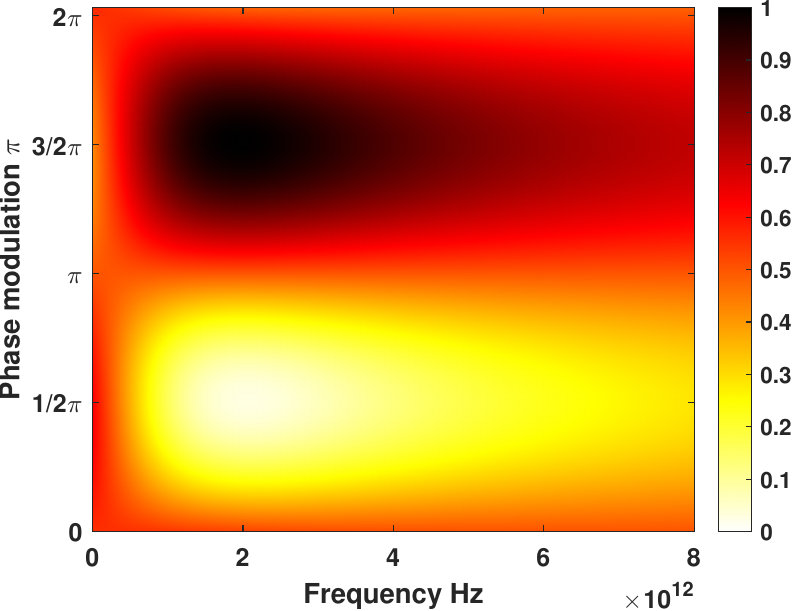}
        \caption{B = 0.6 Tesla.}
    \end{subfigure}
\caption{(a) Absorption of LHC polarizations under sub-tesla magnetic field $B=0.3$ Tesla for the chemical potential $\mu= 0.5 \ eV$ at $T = 10 \ K$ with magneto-CPA frequency 1.859 THz (b) Absorption for magnetic field $B=0.6$ Tesla with magneto-CPA frequency 1.954 THz. Here, the optimized phase modulation is $\Delta\phi= 1.509 \pi$. }
\label{Figure_6}
\end{figure}

	\end{document}